# The double dealing of cyclin D1

Guergana Tchakarska[1] and Brigitte Sola[2,*]

1. Department of Human Genetics, McGill University, Montreal, Quebec, Canada; McGill University Health Centre, Montreal, Quebec, Canada

2. Normandie Univ, INSERM U1245, Unicaen, Caen, France

* Corresponding author. Brigitte Sola: MICAH team, UFR Santé, CHU Côte de Nacre, 14032 Caen Cedex, France. E-mail: brigitte.sola@unicaen.fr. ORCID: 0000-0001-7278-9651

**Abstract**

The cell cycle is tightly regulated by cyclins and their catalytic moieties, the cyclin-dependent kinases (CDKs). Cyclin D1, in association with CDK4/6, acts as a mitogenic sensor and integrates extracellular mitogenic signals and cell cycle progression. When deregulated (overexpressed, accumulated, inappropriately located), cyclin D1 becomes an oncogene and is recognized as a driver of solid tumors and hemopathies. Recent studies on the oncogenic roles of cyclin D1 reported non-canonical functions dependent on the partners of cyclin D1 and its location within tumor cells or tissues. Support for these new functions was provided by various mouse models of oncogenesis. Finally, proteomic and transcriptomic data identified complex cyclin D1 networks. This review focuses on these aspects of cyclin D1 pathophysiology, which may be crucial for targeted therapy.

**Introduction**

The cell cycle and cell proliferation/division are driven by cyclins and their catalytic partners, the cyclin-dependent kinases (CDKs) (Figure 1A). Diverse heterodimeric complexes form, with D-type cyclins (D1, D2 and D3) associating with CDK4 or CDK6 (CDK4/6, two serine/threonine kinases functionally redundant in most tissues, [1]), to control the progression through the G1 phase, the G1-to-S phase transition, thereby initiating DNA replication [2]. The expression, activation, subcellular distribution, stabilization and degradation of D-type cyclins are tightly controlled in normal cells in response to on/off mitogenic signals. In turn, the overexpression, accumulation and mislocalization of D-type cyclins drive human tumors and may confer resistance to chemotherapy. Cyclin D1 is more frequently deregulated than cyclins D2 and D3 in solid cancers and malignant hemopathies [3]. Indeed, the *CCND1* gene encoding cyclin D1 is the second most frequently amplified locus in solid cancers [4]. In hematological malignancies, cyclin D1 is overexpressed due to t(11;14)(q13;q32) translocation, amplification of the *CCND1* gene, deletion or point mutations within the *CCND1* 3'-UTR or the first exon [5, 6]. Cyclin D1 activity is strongly linked to its levels. Tumor cells with high cyclin D1 levels display uncontrolled proliferation in response to a cell cycle dysfunction at the restriction point in the G1 phase (Figure 1A). In addition to this role in regulating the cell cycle, in which it acts in association with CDK4/6 or other partners such as transcription factors, chromatin-modifying enzymes or cytosolic proteins, cyclin D1 also regulates a number of key processes involved in cancer initiation and maintenance. These processes include DNA damage response [7, 8], chromosome duplication and stability [9, 10], senescence [11, 12], autophagy [11, 13], mitochondrial respiration [14, 15], migration [16-18], metabolism [19-21], and immune surveillance [22, 23]. Cyclin D1 shuttles between the nucleus and the cytoplasm and has also been found associated with the outer membrane of the mitochondria [15] or close to the cytoplasmic membranes, at ruffles in migrating cells [24]. Some of the oncogenic functions of cyclin D1 are performed by the nuclear form, whereas others are performed by the cytosolic or membrane-bound forms. Recent reviews on the structure, function, and regulation of cyclin D1 have been published [5, 25-27]. In this review, we focus on the most meaningful findings of recent years. We focus on the pathophysiological functions of cyclin D1 in relation to its subcellular distribution, because these relatively recent observations may have a profound impact on clinical practice. We highlight the importance of addressing the diversity of cyclin D1-associated functions with a view to developing targeted therapies, and the questions that remain unanswered.

**Cyclin D1 shuttles between the nucleus and the cytoplasm in normal cells**

Cyclin D1 levels oscillate and its subcellular distribution changes over the course of the normal cell cycle. Cyclin D1 levels are highly regulated at the transcriptional, post-transcriptional and translational steps [26]. The synthesis of CCND1 mRNA and protein begins when mitogenic stimulation induces quiescent cells to enter G1. Cyclin D1 expression requires the activation of RAS-mediated signaling cascades, which is accompanied by phosphoinositide-3-kinase (PI3K)/AKT-dependent CCND1 translation and a decrease in cyclin D1 degradation. The active cyclin D1/CDK4/6 complexes move to the nucleus, and cyclin D1 levels increase from early to late G1 phase, subsequently decreasing during S phase, when cyclin D1 is exported into the cytoplasm for degradation by the ubiquitin-proteasome system (UPS). This requires the phosphorylation of the threonine 286 (Figure 1B), which is catalyzed by, at least, three kinases: glycogen synthase-3β (GSK3β), p38 MAP kinase (MAPK) and extracellular signal-regulated kinases (ERK1/2) [28, 29]. In the cytoplasm, the ubiquitylation of phosphorylated cyclin D1 is dependent on a SCF complex consisting of the SKP1 adaptor protein, the CUL1 scaffold protein and a F-box protein. At least four E3 ubiquitin ligases of the F-box family ubiquitylate cyclin D1. Three of them (SKP2, FBXW8, FBXO4) are involved in the normal cell cycle whereas as FBXO31 acts in response to genotoxic stresses [30-33]. Mutations or alterations of cyclin D1-specific E3 ubiquitin ligases contribute to cyclin D1 accumulation in cancers [33, 34]. In turn, the promotion of cyclin D1 degradation suppresses cancer cell growth. This can be achieved by the knockdown of *USP2* coding a cyclin D1-specific deubiquitylase or by the use of inhibitors [35, 36].

Cyclin D1 has a nuclear export signal and is a cargo of exportin 1 (XPO1). However, it has no nuclear localization signal and cannot enter the nucleus on its own. Instead, its nuclear entry is favored both by the CIP/KIP (CDK-interacting protein, kinase inhibitory protein) family members and CDK4, and cyclin D1/CDK4/p21$^{CIP1}$/p27$^{KIP1}$ complexes are imported into the nucleus upon activation (Figure 1C).

**Cyclin D1 accumulates in the nucleus and/or in the cytoplasm in tumor cells or specialized tumor regions**

Cyclin D1 is overexpressed in many human solid cancers, including breast cancers, particularly those with the estrogen receptor (ER)-positive subtype [37], head and neck squamous cell carcinoma [38], pancreatic cancer [39], melanoma [40], endometrial cancer [41], colorectal carcinoma [42], and non-small cell lung carcinoma [43] (Figure 2A). In mantle cell lymphoma, a mature B-cell lymphoma, the t(11;14)(q13;q32) translocation leads to the overexpression of cyclin D1 in almost 100% of patients [44]. This translocation is found in 15% of multiple myeloma patients, but cyclin D1 is overexpressed in the absence of a detectable genetic alteration in 50% of multiple myelomas [45].

Cyclin D1 overexpression may occur due to *CCND1* amplification or chromosomal rearrangements, or as a result of impaired degradation of the protein. Mutations and deletions affecting the threonine 286 (Thr286) residue, the site of the glycogen synthase-3β (GSK3β) phosphorylation required for nuclear export by XPO1 and subsequent degradation by the UPS, have been reported in endometrial cancers and esophageal carcinomas [46, 47]. As a result, cyclin D1 is constitutively nuclear in these cells. In MCL cells, genomic deletions and/or points mutations in the 3'-UTR of the CCND1 mRNA result in the generation of a shorter, more stable transcript [48, 49]. Recent genomic studies on MCL identified recurrent mutations in the 5'-coding region of *CCND1* that increased cyclin D1 stability due to a defect in UPS-mediated degradation [6]. In all cases, the mutated proteins preferentially localize to the nucleus. In lung and colorectal cancers, the deubiquitylase USP22 (ubiquitin-specific peptidase 22), a subunit of the Spt-Ada-Gcn5-acetyltransferase (SAGA) transcriptional coactivator, and cyclin D1 levels are correlated. Moreover, by directly deubiquitylating cyclin D1, USP22 protects it from UPS-mediated degradation allowing its accumulation through a novel mechanism [50].

*CCND1* has an A/G polymorphism at nucleotide 870, at the splice donor site of the exon 4-intron 4 boundary. This polymorphism leads to alternative splicing, generating a cyclin D1 variant called cyclin D1b [50] (Figure 1B). The mature cyclin D1b transcript includes exon 4 and intron 4. The corresponding isoform b of the protein lacks the Thr286 required for cyclin D1 degradation. Cyclin D1b is constitutively nuclear [52], and is present in MCL, esophageal, breast, lung and prostate cancers [53]. Cyclin D1b has a lower capacity to phosphorylate the retinoblastoma protein RB1 [52, 54], the central regulator of the cell cycle, but a greater oncogenic capacity than the canonical cyclin D1 (cyclin D1a) [51, 55]. Cyclin D1b (and not cyclin D1a) is capable of transforming NIH-3T3 cells (foci formation) *in vitro* [52]. Cyclin D1b-producing fibroblasts lost contact inhibition, form colonies in soft agar and engraft in immunodeficient mice [54]. These observations highlight the contribution of nuclear functions other than cell cycle regulation to oncogenesis. Moreover, the expression of cyclin D1b, unlike that of cyclin D1a, is unambiguously associated with tumor progression and treatment failure in prostate and breast cancer [53, 56-58]. Using an engineered mouse model in which cyclin D1a was converted into cyclin D1b under the control of the endogenous *CCND1* promoter, Augello and coworkers confirmed that these two isoforms, despite being regulated in a similar fashion, had different functions, with cyclin D1b having a much greater oncogenic potential. Indeed, cyclin D1b fosters transformation of primary fibroblasts and cooperate with the RAS oncogene to drive transformation *in vivo* [56]. Moreover, Gladden and coworkers had previously described a mouse model in which a constitutively nuclear form of cyclin D1 (T286A, Figure 1B) was under the control of the heavy chain immunoglobulin enhancer Eμ [59]. All transgenic mice developed mature B-cell lymphomas, providing further evidence for the oncogenicity of the nuclear form of cyclin D1.

The nuclear accumulation of cyclin D1 is indicative of a high mitotic index in cancer cells, and is considered to be a prognostic marker. On the other hand, high cyclin D1a levels are inversely correlated with with levels of Ki67, a marker of cell proliferation, and are unrelated to clinical outcome [58]. By contrast, cyclin D1b is also upregulated in breast cancers, including ER-negative tumors, independently of cyclin D1a, and is associated with a poor outcome, with recurrence and distant metastasis [58]. In other types of cancer, such as prostate cancer cells, cyclin D1 accumulates in either the cytoplasm or the nucleus. Tumors with cytoplasmic cyclin D1 accumulation have the lowest Ki67 index, whereas those with nuclear cyclin D1 accumulation are of higher grade and have high Ki67 levels [60]. Interestingly, in prostate cancer metastases, a high cytoplasmic cyclin D1 content is predictive of poor outcome [61]. Fusté and coworkers reported that cyclin D1 was mainly cytoplasmic, being found even at the membranes in peripheral and invasive tissues of breast, prostate, colon, and endometrial cancers, suggesting that it might be useful as a marker of invasiveness [62]. Cyclin D1 has been detected in both the nuclear and cytoplasmic compartments of MCL tumor cells. However, it was mostly cytoplasmic in the more aggressive, blastoid form of the disease [63]. In glioblastoma, cyclin D1 is mostly cytoplasmic in evading cells but retained within the nucleus in the tumor mass [64]. These observations confirm that cyclin D1 has unrelated nuclear and cytoplasmic functions.

**Nuclear cyclin D1 has oncogenic roles**

*Nuclear cyclin D1 controls cell proliferation*

In normal cells, cyclin D1 is produced in response to extracellular mitogenic signals and the subsequent activation of RAS-dependent signaling cascades, cyclin D1/CDK4/6 complexes assemble, are stabilized and activated (Figure 1C) [65]. CDK inhibitors of the CIP1/KIP family facilitate cyclin D1/CDK4/6 assembly and nuclear import without inhibiting the kinase activity [64]. Once within the nucleus, D1/CDK4/6 complexes control the G1-to-S phase transition and promote cell cycle progression. Cyclin D1/CDK4/6 complexes phosphorylate RB1 and related pocket proteins (RBL1/2, p107 and p130), thereby dissociating E2F transcription factors, which activate genes controlling the G1-to-S phase transition. When S phase is completed, cyclin D1 is exported to the cytoplasm and degraded. The deregulation of these steps leads to the accumulation of cyclin D1. Cell proliferation becomes independent of extracellular signals, and the cell cycle checkpoints responsible for ensuring genome integrity are bypassed [66].

Cyclin D1/CDK4/6 complexes also phosphorylate several transcription factors, thereby activating or repressing the expression of genes required for cell cycle progression. SMAD3, which inhibits the the anti-proliferative effect of the transforming growth factor (TGF)-β signaling pathway, is a target of cyclin D1/CDK4 complexes [66]. Several cyclin D1/CDK4 phosphorylation sites have been mapped on SMAD3. Mutations of these sites increase SMAD3 transcriptional activity and antiproliferative functions [67]. Cyclin D1/CDK4 phosphorylates and activates the transcription factor forkhead box M1 (FOXM1) thereby maintaining the expression of G1/S-phase genes, promoting the cell cycle and preventing senescence [68].

### *Nuclear cyclin D1 controls transcription*

The best-known function of nuclear cyclin D1 independent of CDK4/6 activity is its ability to act as a cofactor for transcription and to modulate pathways critical for both development, differentiation and tumorigenesis [69]. The DNA binding capacity of cyclin D1, leading to the repression of the *CDKN1A* gene (encoding p21$^{CIP1}$) transcription was reported in 2005 [70]. Genomic and proteomic screens have since revealed that the cyclin D1 interactome comprises more than 30 transcriptional regulators, including nuclear receptors [71]. Mechanistically, the control of transcription involves both transcription factors and chromatin-modifying enzymes.

Among nuclear receptors, cyclin D1 binds and directly controls the transcription of estrogen receptor (ER)α, peroxisome proliferator-activated receptors (PPARs), and androgen receptor (AR) in breast, prostate and liver cancer cells respectively [45]. Cyclin D1 upregulates ERα-mediated transcription. It binds the hormone-binding domain of ERα, thereby increasing ERα-mediated activity even in the absence of the canonical ligand. Cyclin D1 also interacts with steroid receptor coactivators (SRC1 and SRC3), which recruit additional transcriptional cofactors, such as PCAF (p300/CREB binding-associated protein), a histone acetyltransferase, and modify chromatin structure. This increases the transcriptional activity of the ER [72]. The cyclin D1b isoform is present in breast cancer cells but does not bind ERα [73]. By contrast, in prostate cancer cells, cyclin D1a binds AR and represses ligand-dependent activity through two discrete mechanisms: direct binding to prevent the formation of an active conformation and binding to histone deacetylase (HDAC), which mediates transcriptional repression [74]. Cyclin D1b retains the ability to bind AR and stimulates androgen-dependent proliferation [75]. In prostate cancer cells, cyclin D1b promotes the activation of AR-dependent genes associated to a metastatic phenotype, including *SNAI2,* encoding Slug, in particular, independently of an epithelial-mesenchymal transition (EMT) signature [37]. Cyclin D1b and Slug levels are strongly correlated in clinical samples from patients with advanced disease. Cyclin D1 inhibits the transcriptional activity of PPARγ. It also enhances the recruitment of HDAC and histone methyltransferase (such as SuV39H) to the PPAR-response element [76] and inhibits the transcriptional activity of PPARα through an unknown mechanism, thereby acting on fatty acid metabolism and energy homeostasis [20]. The oncogenic function of cyclin D1-mediated PPARs activation remains to be established.

A direct interaction between cyclin D1 and transcription factors, such as STAT3 (signal transducer and activator of transcription 3) and DMP1 (dentin matrix acidic phosphoprotein 1) is involved in cell transformation in breast cancers [77]. In cells with cyclin D1 expression, CDKN1A (encoding p21$^{CIP1}$) is downregulated. Indeed, cyclin D1 is recruited to the p21$^{CIP1}$ promoter by STAT3 and represses its activity [70]. Cyclin D1 bound to DMP1 activates the *ARF* (encoding p14) and *INK4A* (encoding p16) promoters of the *CDKN2A* locus [78]. In normal cells, this activation induces apoptosis or cell cycle delay and protects cells from transformation. In breast cancer cells, DMP1 is frequently lost, and oncogenic signals from *ARF* and *INK4A* promoters are no longer quenched. This observation has been reproduced in mouse models [78]. Cyclin D1 also interacts with other transcription factors from the MYB family. However, the impact of this regulation on cell transformation remains unclear [77].

It has recently been shown that cyclin D1 binds a large number of active promoters in MCL cells, massively downregulating global transcription [79]. It has been suggested that this downregulation of the transcription program has an oncogenic impact on the levels of tumor suppressor proteins, including the products of cell-cycle checkpoint genes. Moreover, the overexpression of cyclin D1 leads to the accumulation of paused polymerase II (Pol II), which may lead to genomic instability by increasing the probability of potential conflicts between DNA replication and transcription machineries. Importantly, cyclin D1 overexpression sensitizes cells to transcription inhibitors, generating a synthetic lethality interaction [79]. This global dysregulation of transcription has not been reported for solid cancers, but the interactomes of cyclin D1 are similar in solid tumors and hemopathies [8].

The DNA-bound form of cyclin D1 regulates gene transcription through histone regulatory enzymes, such as histone deacetylase (HDAC1/3) and histone acetyltransferase (HAT), and chromatin remodeling enzymes such as SUV39H (suppressor of variegation 3-9 homolog) and HP1α (or CBX5, chromobox 5). Further evidence for the control of epigenetic marks by cyclin D1 was recently provided by the finding that cyclin D1 controls global protein methylation by binding directly to G9a lysine methyltransferase [80]. Using ChIP-Seq (chromatin-immunoprecipitation and DNA sequencing) technology to map the global genomic footprint of cyclin D1, Pestell's group identified 3200 DNA regions associated with cyclin D1 throughout the genome. They investigated the transcription sites displaying enrichment in these regions and identified *CTCF*, *SP1*, *ESR1* (encoding ERα), *CREB1* and *HIF1A* genes as master hits [10]. Interestingly, CTCF (or CCCTC-binding factor) is a zinc finger transcription factor serving as an essential anchor site for ERα-mediated proliferation in breast cancer cells [81]. The relevance of the other transcription factors in cyclin D1-mediated tumorigenesis has not been investigated. However, Casimiro and coworkers showed an enrichment for cyclin D1 at genes encoding proteins regulating mitosis and chromosomal stability [10].

***Nuclear cyclin D1 controls the DNA damage response and genomic stability***

DNA replication, metabolic/redox stresses, chemotherapy and ionizing radiation (IR) lead to double-strand DNA breaks (DSBs). Upon DNA damage, DNA-PK (DNA-dependent protein kinase), ATM (ataxia telangiectasia mutant) and ATR (ATM and Rad3-related) apical kinases and their downstream effectors, including the checkpoint kinases (CHK1/2), activate the DDR (DNA damage response) pathway [82]. Depending on the cellular context, the cells then initiate cell cycle arrest and DNA repair. They then pursue the cell cycle or undergo senescence or apoptosis. The notion that the degradation of cyclin D1 is necessary for cells in G1 to repair

DNA and to undergo DNA synthesis dates back to several decades [83]. The expression of the nuclear mutant resistant to UPS-mediated degradation results in greater DNA damage [84]. Cyclin D1 overexpression, as observed in cancer cells, antagonizes the checkpoint-induced cell cycle arrest, disturbs DNA replication and allows damaged DNA to be replicated [7, 84, 85]. Cells that have accumulated DNA breaks are prone to genomic instability, and are characterized by the loss or gain of whole or pieces of chromosomes, chromosomal rearrangements and tumor formation [56, 59, 86, 87]. Casimiro and coworkers showed, in several mouse models of breast cancer, that cyclin D1, independently of CDK4/6, regulated a transcriptional program governing chromosome instability at the level of local chromatin [10, 88]. Cyclin D1 may contribute to oncogenesis by regulating a transcriptional program involved in chromosome instability.

Cyclin D1 directly controls the DDR pathway, as a far downstream effector [89]. Indeed, the silencing of cyclin D1 impairs the DSB repair mediated by the NHEJ (non-homologous end joining) mechanism by inhibiting DNA-PK activation [90]. Moreover, cyclin D1 interacts directly with DNA repair proteins, such as activated ATM/DNA-PK and RAD51 [8, 90]. Cyclin D1 binds to the DNA repair BRCA1 (breast cancer 1) protein, but the consequences of this interaction with regards to DNA repair and cell survival remain unclear [91]. Cyclin D1 binds to RAD51, inducing the phosphorylation of H2AX, the assembly of the RAD51-containing repair machinery, and the recruitment of these repair factors to chromatin [7]. This property is restricted to the isoform a of the protein. Cyclin D1 is also part of the RAD51/BRCA2 (breast cancer 2) complexes at DNA damage sites and has been shown to facilitate HR (homologous recombination)-mediated DNA repair in a CDK4-independent fashion in several types of tumors [8, 92]. Cyclin D1 seems to play a dual role, sustaining the activation of DNA-PK and ATM whilst interacting with the DNA repair machinery. However, it remains unknown how this function of cyclin D1 in DNA repair affects the response of cancer cells to DNA damage *in vivo*.

Steroid hormones, including estrogens and androgens, may induce genotoxic stress, leading in turn to DNA damage. In breast cancer cells stimulated with estrogen, ERα and cyclin D1 assemble at the cytoplasmic membrane, activate the AKT pathway, and induce the formation of γH2AX (the marker of DSBs) foci. Cyclin D1 is recruited to γH2AX foci by RAD51, leading to the recruitment of the DNA machinery responsible for HR-mediated DNA repair [93]. In prostate cancer cells, cyclin D1 promotes DNA repair *via* a similar RAD51-mediated mechanism [13]. The nuclear CDK4/6-dependent and -independent functions of cyclin D1 are represented in Figure 3.

**Cytoplasmic and membrane-associated cyclin D1 regulates tumor cells invasion and dissemination**

Cell migration and invasion are hallmarks of cancer, leading to tumor cell expansion and dissemination through metastasis [94]. Tumors use several strategies to spread within tissues, involving mesenchymal, ameboid single-cell migration or collective movements. Most of cell processes involved in migration (cell polarization and protrusion, cell contraction, cell detachment) require a reorganization of the cytoskeleton [95]. Cyclin D1 has been known to play a major role in tumor cell migration, since the seminal work of Pestell's group on *Ccnd1-/-* mouse macrophages [16] and embryonic fibroblasts [17]. Epithelial cells depleted of cyclin D1 display enhanced adhesion and reduced migration [17, 18]. Cyclin D1 promotes cell motility by inhibiting Rho-activated kinase II (ROCKII) signaling and repressing the thrombospondin 1 (TSP1), a metastasis repressor. This process requires cyclin D1/CDK4/6 activity and the phosphorylation of cytoskeleton proteins involved in cell shape remodeling (Table 1) [96]. Zhong and coworkers identified several cytoskeleton-related proteins phosphorylated by cyclin D1/CDK4/6 complexes [24]. Cytoplasmic cyclin D1 regulates cell invasion and metastasis through the phosphorylation of paxillin and RAC1 activation, and promotes cell ruffling and invasion [97]. Cyclin D1 has been found associated with paxillin at the cell membrane in those migrating cells. Paxillin is also the downstream target of cytoplasmic cyclin D1/CDK4/6 In glioblastoma [64]. Clinical observations support this major role of cytoplasmic cyclin D1 in invasion and metastasis [62, 64].

In B lymphoma tumor cells, a high cytoplasmic cyclin D1 content is associated with a greater invasion capacity. Cytoplasmic cyclin D1 associates with structural proteins of the cytoskeleton, affecting cell shape. Cytoplasmic cyclin D1 is present in the blastoid invasive variant of MCL, associated with a poor prognosis [63]. Indeed, MCL cells can engraft at various locations in the body, according to their cytoplasmic cyclin D1 content and invasive capacity [63].

In addition to transcriptional regulation, cyclin D1 regulates non-coding miRNAs [96] and induces the expression of DICER, the central regulator of miRNA maturation at transcriptional level [99]. The cyclin D1-mediated proliferation and migration of breast cancer cells are DICER-dependent.

The nuclear accumulation of cyclin D1 is a marker of proliferation. If cytoplasmic cyclin D1 levels are a *bona fide* marker of invasiveness, then it will be important to determine the distribution of cyclin D1 in tumors. Another important finding, albeit not replicated in all settings (Table 2), is the CDK4/6-dependence of cyclin D1-mediated migration and invasion. The targeting of CDK4 and CDK6 is therefore highly relevant for therapeutic purposes [2, 27].

**Cell metabolism is regulated by both nuclear and cytoplasmic cyclin D1**

Studies of transgenic mice expressing an antisense construct against cyclin D1 (ErbB2-cyclin D1 antisense) or overexpressing cyclin D1 (MMTV-cyclin D1) in the mammary gland have shown that cyclin D1 inhibits oxidative glycolysis, lipogenesis and mitochondrial activity [14]. Cyclin D1 decreases hexokinase 2 (HK2) abundance and transcription; HK2 is the first enzyme of the glycolysis pathway [14]. Thus, cyclin D1 depletion from normal or cancerous breast cells leads to an increase in the levels of the glycolytic enzyme pyruvate kinase (PK) and of the lipogenic enzymes acetyl-CoA carboxylase (ACC) and fatty acid synthase (FASN). In B lymphocytes, HK2 and cyclin D1 compete for binding to VDAC (voltage-dependent anion channel) at the outer membrane of the mitochondria [15]. Cyclin D1 bound to VDAC releases HK2, which may then participate in and enhance glycolysis (Caillot *et al*., in preparation). In hepatocytes, cyclin D1 represses gluconeogenesis and oxidative phosphorylation by inhibiting the PPARγ co-activator, PGC1α. This inhibition is CDK4/6-dependent and is dependent on the fasting and refeeding of hepatocytes [100]. It has been suggested that this regulatory pathway has an effect on diabetes, rather than cancer. Again in hepatocytes, cyclin D1 inhibits the glucose-mediated induction of lipogenic genes by repressing the carbohydrate response element binding protein (ChREBP) and hepatocyte nuclear factor 4α (HNF4α), which are important regulators of glucose sensing and lipid metabolism [101]. CDK4 activity is required for the inhibition of ChREBP-mediated transcription but unnecessary for HNF4α-regulated lipogenesis. Cyclin D1 thus inhibits both glycolysis and lipogenesis, *via* CDK4-dependent and -independent mechanisms.

Moreover, cyclin D1 controls mitochondrial biogenesis and functions by inhibiting the nuclear respiratory factor 1 (NRF1). This reduction in mitochondrial activity requires CDK4 [102]. In B cells, including MM tumor cells, cyclin D1 expression is also linked to a decrease in mitochondrial activity. The binding of cyclin D1 to VDAC impairs ADP access to the inner mitochondrial membrane [15]. An analysis of cyclin D1 interactors in various cancer cells identified several proteins involved in metabolism, such as fatty acid synthase, a lipogenic enzyme, glyceraldehyde-6-phosphodehydrogenase, pyruvate kinase M1/M2, and enolase 1, all involved in glycolysis [8, 69 and Caillot *et al*., in preparation]. Interestingly, Wang and coworkers reported that cyclin D3/CDK6 complexes play an unique role in glucose metabolism, by phosphorylating and inhibiting two keys enzymes: pyruvate kinase M2 and 6-phosphofructokinase. This inhibition redirects glycolytic intermediates into the pentose phosphate and serine pathways and sustains prosurvival functions [103]. Cyclin D1/CDK4/6 complexes may have a similar oncogenic function.

**Animal models confirm the oncogenic functions of cyclin D1 *in vivo***

Gain- and loss-of-function studies in transgenic mice have confirmed the diverse functions of cyclin D1, including, in particular, its crucial role in embryogenesis and development (not discussed here). The overexpression of cyclin D1 or the expression of the b isoform of this protein revealed oncogenic functions. Indeed, in MMTV-cyclin D1 transgenic mice in which cyclin D1 is under the control of the mammary tumor virus (MMTV) promoter, mammary adenocarcinomas developed in mammary tissues [104]. Cyclin D1 has also been shown to be essential for breast cancer maintenance and progression *in vivo*. Indeed, cyclin D1-deficient mice are resistant to MMTV-ErbB2- and MMTV-Ras-induced breast cancers [105]. ErbB2-driven mammary carcinomas display arrest and senescence following cyclin D1 knock-down or the inhibition of CDK4/6 [106]. The kinase activity of CDK4/6 seems to be required for breast tumorigenesis, because mice expressing a kinase-deficient mutant form of cyclin D1 (CCND1 K112E) are protected against ErbB2-induced adenocarcinoma [107]. Cyclin D1 acts as a mediator of mammary tumorigenesis induced by ERBB2 in a CDK-dependent manner [108].

Transgenic mice expressing *CCND1* under the control of thee Eμ enhancer (mimicking what occurs in B-cell lymphoma) do not develop detectable lymphoproliferation [109, 110]. In sharp contrast, equivalent constructs with the nuclear mutant *CCND1* T286A, rather than the wild-type gene, induce a mature B-cell lymphoma [59]. Transgenic mice expressing the same T286A mutant under the control of the MMTV promoter develop mammary carcinomas earlier and at a higher frequency than mice transgenic for MMTV-*CCND1* mice [111]. Thus, the transformation process may follow two different courses, depending on *CCND1* status: if *CCND1* is overexpressed as a consequence of oncogenic signals, such as RAS, ERBB2 but also STATs, Notch, or NF-kB (not documented here), then transformation is dependent on CDK activity, whereas, if *CCND1* is overexpressed as a function of mutation or gene rearrangement, cyclin D1 is the primary driver for oncogenesis. Hepatocarcinomas, colon carcinomas and skin papillomas have been shown to result from the overexpression of cyclin D1, in the appropriate transgenic mouse models [27].

The notion that the chromosome instability driven by the nuclear cyclin D1 isoform is part of the oncogenic process, is also supported by results of *in vivo* experiments. In the Eμ-*CCND1* T285A transgenic mouse model, malignant lymphocytes display aneuploidy and chromosomal translocation due to alterations in DNA replication [84]. Furthermore, an enrichment in the expression of genes associated with chromosome instability is observed in breast cancer cells in an inducible model of acute cyclin D1 expression in mammary tissue, and in the chronic MMTV-*CCND1* expression model [10]. In this second transgenic model, the kinase-dead mutant (K112E) induces mammary carcinoma, and tumor cells have high chromosome instability scores [88]. However, it remains unclear how cyclin D1 regulates the chromosome instability-associated genes signature *in vivo*. It is possible that cyclin D1 overexpression leads to a conflict between transcription and replication machineries [79].

However, if the downregulation of the global transcriptome, described in B lymphoma cells, is a hallmark of cyclin D1-expressing tumors, then an association of transcription inhibitors with CDK inhibitors may significantly improve tumor burden.

**Open questions**

The two major functions of cyclin D1 associated with tumorigenesis (nuclear functions in both cases) are cell cycle regulation and chromosome stability control. Cell cycle control is dependent on cyclin D1/CDK4/6 complexes and their kinase activity, whereas chromosome stability control is CDK4/6-independent. As for the dysregulation of the cell cycle, the induction of chromosome instability is mediated by nuclear cyclin D1. Does the nuclear cyclin D1 consist entirely of isoform b or, does it consist of accumulated and mutated isoform a? Both isoforms are produced under the control of the same promoter (Figure 1B) and they are expressed in concert. However, although cyclin D1b is known to be present in several tumor types, it is not clear whether it is systematically produced in these tumors. It is essential to resolve this issue, given the oncogenic functions of cyclin D1b. Systematic studies of tumor cells with adequate tools for distinguishing between the two isoforms are required.

The polymorphism (G870A) at the exon 4 splice donor site in *CCND1* is responsible for the alternative splicing leading to the generation of the b isoform (Figure 1B). The allele with a G at nucleotide 870 (codon 242) may preferentially encode the CCND1a transcript whereas the allele with an A residue at the same position may encode the b isoform [53]. Nevertheless, cyclin D1b is found in tumors homozygous for the G allele in a mouse model of rectal carcinogenesis [112]. Moreover, meta-analyses on tumor tissues have yielded inconclusive results about the relationship between the A/G polymorphism and tumor initiation and/or progression. Two groups have independently characterized RNA binding factors that might control cyclin D1b production [113, 114]. In prostate cancer cells, SRSF1 (serine and arginine-rich splicing factor 1) preferentially associates with and promotes the accumulation of the isoform b transcript from the G allele. SRSF1 is induced during the progression of prostate cancer [1131]. In sharp contrast, in the same tumor type, Sam98 directly affects the *CCND1* splicing through a preference for the A allele. Sam98 expression is correlated with cyclin D1b accumulation in clinical samples [114]. Interestingly, the catalytic subunit Brm (Brahma) of the SWI/SNIF complex involved in chromatin remodeling, described as a regulator of *CCND1* alternative splicing, associates with Sam98 [115]. Several mechanisms impairing cyclin D1b production have been identified, but their effects on malignant transformation remain to be established.

Further insight has been obtained from high-throughput analyses of the cyclin D1 transcriptome and interactome in tumor cells [8, 63, 69]. More than 100 proteins interacting with cyclin D1 have been identified in tumors in different proteomics screens [8, 63, 116]. Cyclin D1 interactors are involved in cell cycle regulation, DNA repair, DNA replication, metabolic reprogramming, and apoptosis (Table 2). Are these functions physiological or associated with gains of function occurring as a consequence of cyclin D1 overexpression? It will be important to explore these points in the context of the development of anti-CDK4/6 therapies.

Selective inhibitors of CDK4/6 (CDK4/6i, including palbociclib, the first used in clinic) are currently in clinical trials (https://clinicaltrials.gov/) and showed positive outcomes, especially for ER-positive breast cancers and B-cell lymphomas [2]. CDK4/6i prevent RB1 phosphorylation and cause cell cycle arrest in G1. In turn, CDK4/6i show efficacy only if RB1 is functional in tumor cells. However, hyperactivation of cyclin E/CDK2 complex which substitutes for CDK4/6 and phosphorylate RB1 is one major mechanism of CDK4/6i resistance.

This could be due to high levels of CDK2, amplification of *CCNE1* (coding for cyclin E), loss of p21$^{CIP1}$ or p27$^{KIP1}$ [117]. Other than RB1, FOXM1 and SMAD3, both involved in cell proliferation, are phosphorylated by cyclin D1/CDK4/6. To our knowledge, the effects of CDK4/6i on TGF-β-mediated cell proliferation has not been assessed. Among the CDK4/6 substrates are cytoskeleton structural proteins and proteins regulating adhesion, migration, and invasion (Table 2). Whether CDK4/6i affect these proteins and non-canonical functions of cyclin D1 in RB1-negative and RB1-positive tumors remains to be established. In any case, CDK4/6i may contribute to a therapeutic effect. It would be of great interest to investigate the combination of CDK4/6i with already characterized signaling pathways inhibitors.

**Conclusions**

Cyclins D1, D2 and D3 have similar cell cycle regulatory functions. However, only cyclin D1 is significantly overexpressed of in solid cancers and hemopathies and has CDK4/6- and RB1-independent functions promoting cancer development. Several of these functions - the transcription regulation, chromosome stability and migration control - are well-documented and known to be involved in tumorigenesis. Others, such as the regulation of metabolism and the non-coding genome, are beginning to emerge, opening up to new perspectives for cancer management, but also rendering our vision more complex.

**Acknowledgments**

We thank the present and former members of the MICAH team involved in cyclin D1 research for their invaluable contributions.

**Funding**

This work was supported by a grant from the *Ligue contre le Cancer* (CD76) awarded to BS.

**Disclosure statement**

The authors have no conflict of interest to declare.

**Figure legends**

**Figure 1**. Structure and function of cyclins and cyclin D1
**A**. Schematic representation of the cell cycle. Following mitogenic signals, eukaryotic cells exit quiescence and irreversibly enter in G1 phase after the restriction (R) point. The successive G1, S, G2 and M phases are controlled by cyclins and their cognate CDKs, as indicated. **B**. Structure of the *CCND1* gene, mRNAs and cyclin D1 proteins. The *CCND1* gene (NG-007375.1) comprises five exons (E1-5) separated by four introns (I1-4). It encodes the full-length canonical cyclin D1 product (cyclin D1a) of 295 amino acids (aa). Through alternative splicing, *CCND1* generates two types of mRNA: the canonical form (NM_053056.2) and the so-called "b" form, which includes the intron 4 encoding an additional stretch of 33 amino acids. The corresponding protein isoforms, "a" and "b", are identical over the first 240 amino acids from the N-terminus, but have different C-termini. Isoform "b" lacks the threonine 285 residue and the PEST sequence (aa 241-290) required for degradation, and the LxxLL motif (aa 251-257) required for ligand-dependent interactions with nuclear receptors. By contrast, both isoforms contain the cyclin box required for CDK4/6 and CIP1/KIP1 family binding, and the LxCxE motif (aa 5-9) required for RB1 binding. **C**. Schematic representation of the G1-to-S phase transition. For cells to exit quiescence (G0), they require mitogenic signals that activate the RAS signaling pathways and, to a lesser extent, the Wnt/β-catenin and NF-κB pathways. Cyclin D1 is activated at several levels after its translation: stability, assembly with its CDK4/6 partners, and import into the nucleus *via* the CKIs of the CIP/KIP family. Cyclin D1/CDK4/6 complexes accumulate during the G1 phase, until the start of DNA replication. Cyclin D1 is then exported to the cytoplasm, where it is degraded by UPS. Cyclin D1/CDK4/6 phosphorylates and inactivates RB1 (and pocket proteins RBL1/2, p107 and p130) facilitating the dissociation of E2F transcription factors and the activation of the necessary genes for DNA synthesis. Cyclin D1 degradation is required for the progression of S phase. The GSK3β kinase phosphorylates cyclin D1, which is then taken up by XPO1 and exits the nucleus. Specific E3 ubiquitin ligases ubiquitinylate cyclin D1, which is then degraded by the proteasome machinery [65].

**Figure 2**. Deregulation of *CCND1* in human cancers
**A**. The frequencies of mutation, amplification, and deletion in human cancers are depicted in the graph. Data were obtained from the cBioPortal for Cancer Genomics (//www.cbioportal.org/). We analyzed 10,259 samples from 9,682 patients with 25 different cancers from the Cancer Genome Atlas. **B**. The types and frequencies of missense, in-frame and truncating mutations of *CCND1* observed with regards to the cyclin D1 protein structure are shown.

**Figure 3**. Nuclear functions of cyclin D1
Cyclin D1 activity associated with CDK4/6 is the canonical function of this protein in the control of the cell cycle and cell proliferation. Once activated, cyclin D1/CDK4/6 complexes phosphorylate RB1 (and RBL1/2), leading to their inactivation and the release of E2F family members. The transcriptional program controlled by E2F includes a battery of genes required for G1-to-S phase transition [45]. In addition to its role in cell cycle regulation independent of kinase activity, cyclin D1 acts in complexes with CDK4/6, to sequester the CKIs p21*CIP1* and p27*KIP1,* thereby indirectly controlling the activity of cyclin E/CDK2 complexes acting during the transition G1-to-S. Cyclin D1/CDK4/6 complexes phosphorylate SMAD3 and downregulate the transcription of genes involved in growth inhibition from TGF-β family [67].

Cyclin D1/CDK4 complexes phosphorylate the MEP50 cofactor for PRMT5, an arginine methyltransferase controlling methylation and transcriptional repression. In a mouse model of B-cell lymphomagenesis, Aggarwal and coworkers showed that nuclear cyclin D1 triggered an increase in MEP50/PRMT5 activity, decreasing the activity of CUL4, the E3 ligase of CDT1, the replication licensing protein. Moreover, PRMT5 appears to be necessary for cyclin D1-mediated transformation [118]. However, unlike the targeting of RB1, SMAD3 and FOXM1, this seems to be important during the S phase. We did not comment on this point in the text, but we recommend another recent review including this aspect [77]. Forkhead box M1 (FOXM1) is a critical target of cyclin D1/CDK4/6. Once stabilized by phosphorylation, FOXM1 maintains G1/S phase expression, and protects cancer cells from senescence [68]. FOXM1 appears to be an oncogenic driver for solid tumors [119].

Independently of CDK4/6, cyclin D1 upregulates (ERα) or downregulates (AR, PPARs) nuclear receptor-mediated transcription by binding directly to the receptors and their co-activators (SRC1, AIB1). This regulation may occur directly or through the recruitment of chromatin modifiers of the histone deacetylase or acetyltransferase families (HDAC, HAT) and SuV39H, HP1α.

Finally, cyclin D1 binds to proteins of the DNA repair machinery (BRCA1/2, RAD51) and participates to both NHEJ and HR mechanisms of DNA repair.

**Table 1.** Cytoplasmic functions of cyclin D1 in adhesion, migration and invasion

| Cell type | CDK4/6 activity | Target | Cytoplasmic form | Function | Reference |
|---|---|---|---|---|---|
| Fibroblast | Yes | ROCK2/Cofilin TSP1 | not reported | Adhesion Migration | [120] |
| Mammary epithelial cell* | No | p27$^{KIP1}$ | Yes | Migration | [17] |
| Mammary epithelial cell | Yes | RHOA/ROCK2 | not reported | Migration | [18] |
| Breast cancer | Yes | Filamin A | not reported | Migration Invasion | [24] |
| Prostate** Fibroblast | not reported | PACSIN 2 | YES | Migration | [121] |
| Prostate cancer Fibroblast | Yes | Paxillin/RAC1 | not reported | Membrane ruffling Invasion | [62] |
| Multiple myeloma | not reported | F-actin fiber ERK1/2 signaling | not reported | Adhesion | [122] |
| Mantle cell lymphoma | not reported | Cytoskeleton proteins | Yes | Migration Invasiveness | [63] |
| Glioblastoma | Yes | Paxillin/RAC1 RALA | Yes | Tumor dissemination | [64] |

* In this report, it was shown that the nuclear cyclin D1b form failed to bind p27$^{KIP1}$, to inhibit RHOA/ROCK activity and had no effect on cell migration. ** In this report, cyclin D1b failed to bind PACSIN B and to repress cell migration. Abbreviations: ERK1/2, extracellular signal-regulated kinase 1/2 or MAPK1/2; KIP1, kinase inhibitory protein or CDKN1B, cyclin-dependent kinase inhibitor 1B; RAC1, RAC family small GTPase 1; ROCK, RHO-associated coiled coiled containing protein kinase; TSP1, thrombospondin 1.

**Table 2**. Pathway enrichment analysis of cyclin D1 partners in human cancer cell lines

| Name | WikiPathway | Adj. *p*-value | Odds ratio | Score |
| --- | --- | --- | --- | --- |
| DNA damage response | WP707 | 1.88e-12 | 24.99 | 828.50 |
| miRNA regulation of DNA damage response | WP1530 | 1.14e-12 | 23.93 | 779.21 |
| Retinoblastoma gene in cancer | WP2446 | 1.01e-12 | 21.04 | 695.86 |
| G1-to-S cell cycle control | WP45 | 1.81e-10 | 22.47 | 606.18 |
| DNA mismatch repair | WP531 | 7.65e-4 | 43.57 | 446.26 |
| Cell cycle | WP179 | 5.06e-11 | 15.25 | 434.30 |
| miRNAs involved in DNA damage response | WP1545 | 1.24e-4 | 34.86 | 431.49 |
| Integrated cancer pathway | WP19718 | 1.24e-4 | 20.80 | 354.06 |
| Signaling pathway in glioblastoma | WP2261 | 5.36e-8 | 15.94 | 336.47 |
| Valproic acid pathway | WP3871 | 2.15e-3 | 30.17 | 272.67 |
| DNA replication | WP466 | 3.26e-5 | 18.67 | 263.14 |
| DSBs and cellular response via ATM | WP3959 | 9.40e-6 | 16.64 | 256.68 |
| Rb-E2F1 signaling | WP3969 | 2.62e-3 | 28.01 | 246.60 |
| FSGS | WP2572 | 4.55e-6 | 14.52 | 237.87 |
| DSBs and cellular response via ATR | WP4016 | 9.31e-6 | 13.07 | 203.19 |
| PPARα pathway | WP2878 | 8.49e-4 | 20.11 | 202.12 |
| Apoptosis-related network (Notch) | WP2864 | 1.12e-4 | 14.80 | 187.71 |
| ATM signaling pathway | WP2516 | 3.41e-4 | 16.34 | 183.82 |
| Metabolic reprogramming | WP4290 | 4.13e-4 | 15.56 | 171.26 |
| IL-1 induced NF-κB activation | WP3656 | 2.70e-2 | 26.14 | 156.50 |

The 153 proteins identified as cyclin D1 interactors in at least one of the five cancer cell line tested by Sicinski's group (mantle cell lymphoma, breast cancers, pharyngeal cancer, colon cancer, [8]) were analyzed with the Enrichr tool [123] for enriched WikiPathways [124] The top 20 enriched pathways are indicated.

Abbreviations: Adj., adjusted; ATM, ataxia telangiectasia mutated; ATR; ataxia telangectasia and Rad3 related; DSB, double-strand break; FSGS, primary focal glomerulosclerosis; IL, interleukin; NF-κB, nuclear factor κB; PPAR, peroxisome proliferator-activated

Figure 1

Figure 2

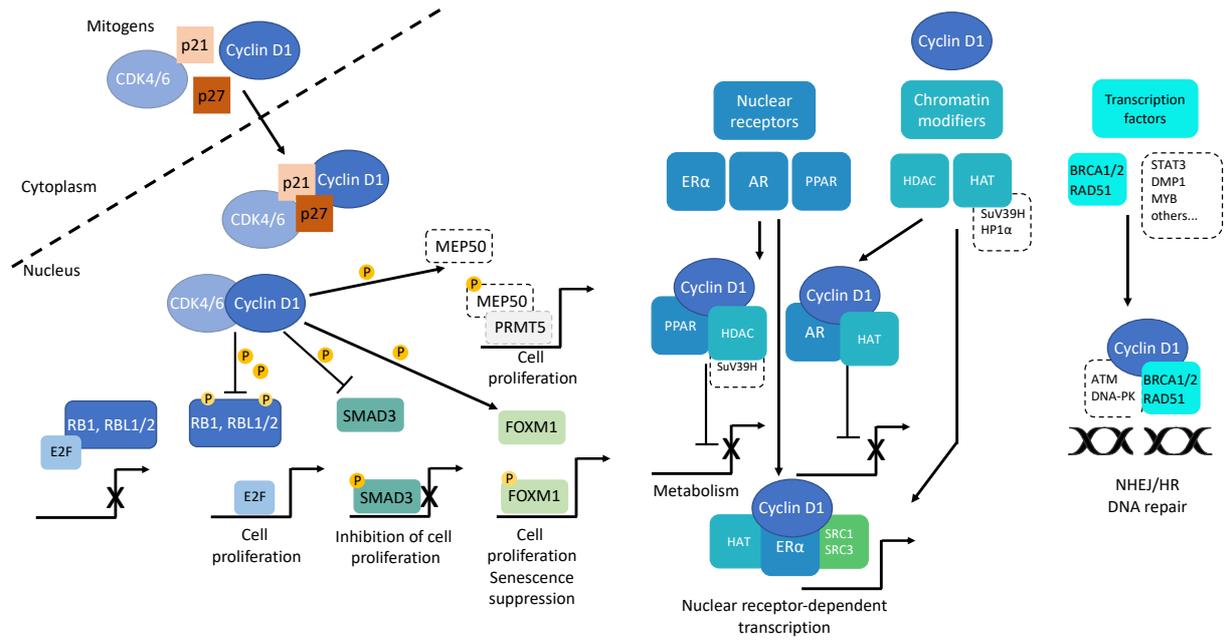

Figure 3